\documentclass[final,5p,times,twocolumn]{elsarticle}

\usepackage{amssymb}

\journal{Physica C}

\begin{document}

\begin{frontmatter}



\title{Puzzling evidence for surface superconductivity in the layered dichalcogenide Cu$_{10\%}$TiSe$_2$}


\author[label1,label2]{F.~Levy-Bertrand}
\author[label1,label2]{B.~Michon}
\author[label1,label2]{J.~Marcus}
\author[label1,label3]{C.~Marcenat}
\author[label4]{J.~Ka\v{c}mar\v{c}\'{i}k}
\author[label1,label2]{T.~Klein}
\author[label1,label2]{H.~Cercellier}

\address[label1]{Univ. Grenoble Alpes, Inst NEEL, F-38000 Grenoble, France}
\address[label2]{CNRS, Inst NEEL, F-38000 Grenoble, France}
\address[label3]{CEA, INAC-SPSMS, F-38000 Grenoble, France}
\address[label4]{Centre of Low Temperature Physics, Institute of Experimental Physics, Slovak Academy of Sciences, 04001 Ko\v{s}ice, Slovakia}

\begin{abstract}
We report on specific heat and magnetotransport measurements performed on superconducting Cu$_{10\%}$TiSe$_2$ single crystals. We show that superconductivity persists in transport measurements up to magnetic fields $H_R$ well above the upper critical field $H_{c2}$ deduced from the calorimetric measurements. Surprisingly this "surface" superconductivity is present for all magnetic field orientations, either parallel or perpendicular to the layers. For $H\|ab$, the temperature dependence of the $H_R/H_{c2}$ ratio  can be well reproduced by solving the Ginzburg-Landau equations in presence of a surface layer with reduced superconducting properties. Unexpectedly this temperature dependence does not depend on the field orientation.
\end{abstract}


\end{frontmatter}


\section{Introduction}

It has been shown several decades ago that surface superconductivity can persist up to temperatures much larger than the {\it bulk} critical temperature in the presence of a magnetic field parallel to the sample surface \cite{deGennes}. Beyond this standard effect, surface and/or interface superconductivity is at work in various systems leading to intriguing properties. For instance the presence of misfit dislocations in semiconducting monochalcogenide heterostructures result into the formation of inversion layers that turn out to be superconducting with unexpectedly high critical temperatures \cite{Fogel}. Beside, thanks to crystal growth progress, artificial layered metamaterials or single layer are nowadays accessible, generating fascinating behaviors (for a review see ref.[3]). For example, the interface between the LaAlO$_3$ and SrTiO$_3$ insulators happens to be superconducting\cite{LaAlO3/SrTiO3} or the single FeSe epitaxial layer becomes superconducting at a critical temperature rising up to about 100K, exceeding by about an order of magnitude the bulk value\cite{FeSe/SrTiO3}. 

Surface effects might also play a fundamental role in unconventional superconductors. Indeed, those materials  {\it naturally} present layered structures.  It has been suggested that a new type of superconductivity, called pair density wave \cite{Berg}  could for instance exist in those Óstripped superconductorsÓ consisting of an array of superconducting regions separated by (narrow) insulating regions. In those systems, superconductivity also often competes with another electronic and/or magnetic instability and spatial modulations of the order parameter of this competing phase may then lead to interfacial (or surface) ÓhiddenÓ superconductivity with enhanced critical temperature\cite{Moor}. 

 In this paper we show some evidence for the presence of some {\it puzzling} surface superconductivity in the  Cu$_{10\%}$TiSe$_2$ dichalcogenide.  1T-TiSe$_2$ is an octahedrally coordinated dichalcogenide which exhibits a charge density wave (CDW) accompanied by a periodic lattice distortion below 200~K. Recent discoveries of chirality\cite{Ishioka} and ultrafast dynamics of charge carriers in the charge-ordered state \cite{Rossnagel}, as well as a strong theoretical activity have led to a unified picture in which the CDW transition is driven by a cooperative exciton-phonon mechanism \cite{vanWezel, Castellan, Zenker}. Upon doping by Cu intercalation, the CDW is continuously suppressed and a superconducting dome emerges \cite{Morosan2006}. Superconductivity reaches its maximum critical temperature when the CDW is close to being suppressed (or at least is too faint to be detected). We present here a combined study of the transport and specific heat properties of high quality, optimally doped single crystals. We show that the resistivity deviates from its normal  state value for fields much larger than the upper critical field $H_{c2}$ corresponding to the establishment of {\it bulk} superconductivity, as deduced from specific heat measurements.  Surprisingly this {\it surface} superconductivity is only weakly dependent on the orientation of the applied magnetic field and remains clearly visible even for magnetic fields perpendicular to the sample surface.

\begin{figure}[h]
\begin{center}
\resizebox{8.5cm}{!}{\includegraphics{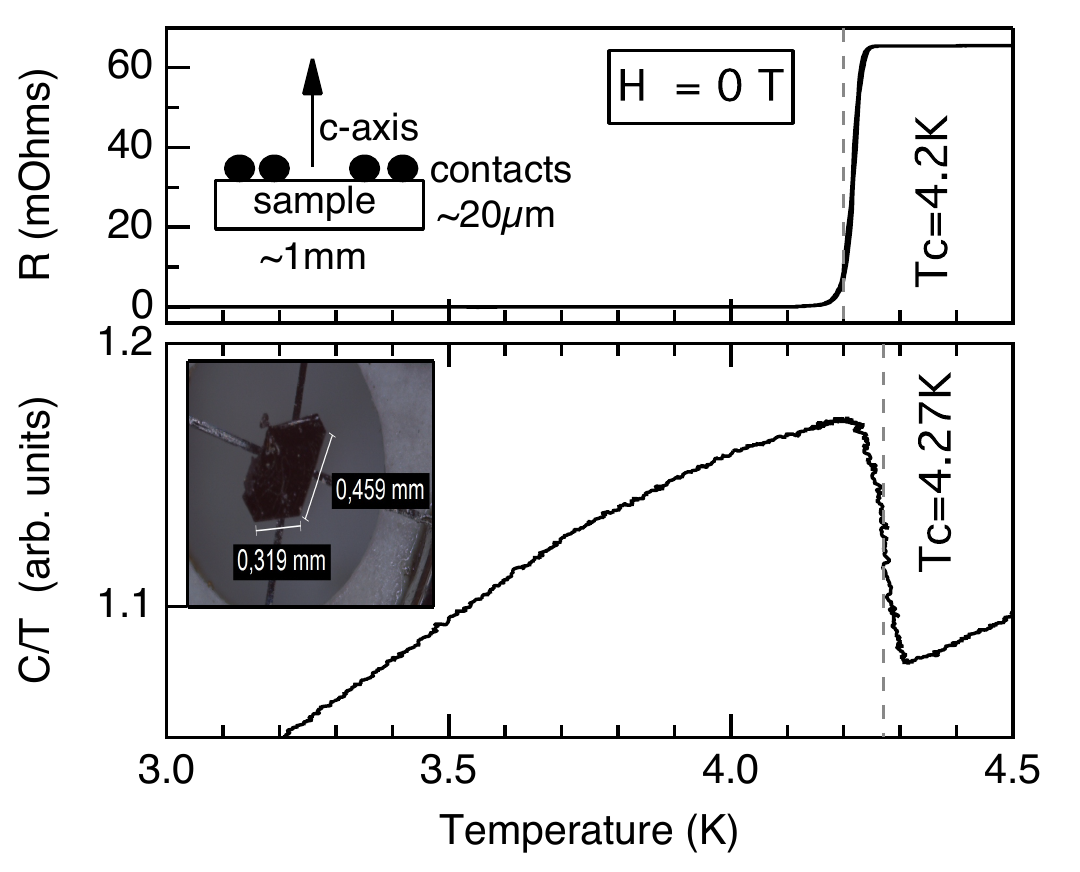}}
\caption{Temperature dependence of the resistivity (upper panel) and specific heat (lower panel) in Cu$_{10\%}$TiSe$_2$ single crystals. As shown, very sharp anomalies are observed in both measurements and the small shift in the critical temperature most probably reflects the small dispersion of the critical temperature within the sample batch. The inset of the upper panel sketches the contact geometry and the photo in the lower panel displays the sample mounted on the specific heat probe.}
\label{fig1}
\end{center}
\end{figure}

\section{Samples and experiments}

Single crystals of Cu$_{x}$TiSe$_2$ were grown by mineralization. Stoichiometric amounts of Cu, Ti and Se powders were heated up to 650$^o$C in a quartz tube under vacuum. Sub-milimetric to milimetric hexagonal platelets, with thickness on the order of 20~$\mu$m, were obtained (see inset of figure~\ref{fig1}, lower panel). X-ray diffraction confirms that the samples crystallized within the P-3m1 space group. Slight unfolding Bragg peaks were observed, revealing stacking faults along the $c$-axis perpendicular to the hexagonal plane. Scanning electron-microscopy has been performed on some of the crystals confirming that the copper doping is homogenous within $\pm$0.4$\%$ (but smaller than the 10$÷\%$ nominal value by $\sim 1-2\%$).

AC specific heat and four-probe resistivity measurements  have been performed on several crystals. The magnetic field orientation has been varied from the layer $ab$-plane, to the axis perpendicular to the hexagonal structure (e.g. $c$-axis). In transport measurements, four electrical contacts were attached on the sample surface (see sketch in the inset of the upper panel of Fig.1) using silver paint and the current was applied in the layer plane, roughly along the $a$-axis. $C_p$ measurements have been performed using a high sensitivity AC technique in which heat was supplied to the sample by a light emitting diode and the induced temperature oscillations were recorded with a thermocouple\cite{kavcmarvcik2010specific}. Special care has been taken to reduce the electrical noise down to $\sim 5$ pV (at 0.5 K) using cold transformers and very low noise home built pre-amplifiers.  

\section{Results and discussion}

 Figure~\ref{fig1} displays the zero magnetic field temperature dependence of the resistivity (upper panel) and specific heat (lower panel). As shown a very well resolved specific heat jump is observed at the critical temperature $T_c = 4.27 \pm 0.05$ K (the error bar corresponds to the width of the jump). This transition is accompanied by a sharp drop of the resistivity to zero (see upper panel). The slightly lower corresponding $T_c$ is most probably reflecting the dispersion of the critical temperatures within the different crystals of the batch.  Note that the samples studied here exhibit the highest $T_c$ with the smallest transition width ever reported in this system. 

\begin{figure}[h]
\begin{center}
\resizebox{8.5cm}{!}{\includegraphics{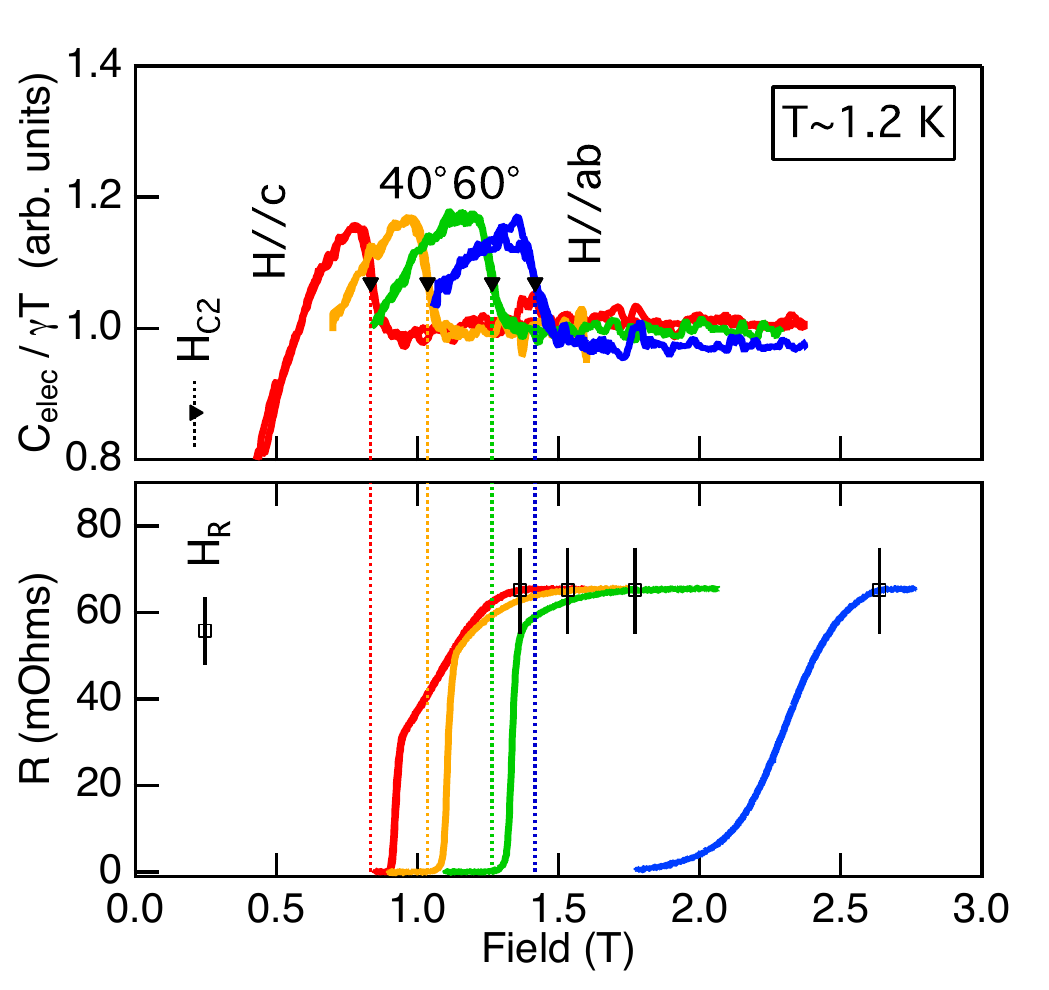}}
\caption{Magnetic field dependence of the resistivity (lower panel) and specific heat (upper panel) at $T$~=~1.2~K for the indicated magnetic field orientation. Vertical dotted lines corresponds to the mid point of the specific heat anomaly and hence to the upper critical field $H_{c2}$ corresponding to the onset of {\it bulk} superconductivity. As shown, the resistivity clearly deviates from its normal state value for fields $H_R$ much larger than $H_{c2}$ for all orientations of the magnetic fields. }
\label{fig2}
\end{center}
\end{figure}

Figure~\ref{fig2} displays the field dependence of resistivity (lower panel) and specific heat (upper panel) at $T=1.2$ K (as an example) for the indicated magnetic field orientations. As shown, the shape of the $C_p$ anomaly remains unchanged whatever the field orientation, hence allowing an unambiguous determination of the angular dependence of the upper critical field $H_{c2}$ (dashed vertical lines in Fig.2). Note that the clear observation of a well defined specific heat jump is another indication for the high quality of the crystals.  As expected for anisotropic superconductors, the superconducting transition is shifted towards higher fields as the applied magnetic field is moved from the $c$-axis to the $ab$-plane and this angular dependence can then be well fitted using the standard London model for (3D) anisotropic superconductors :
\begin{equation}
H_{c2}(\theta)=\left[\left[\frac{sin \theta}{H_{c2//ab}}\right]^2+\left[\frac{cos \theta}{H_{c2//c}}\right]^2\right]^{-1/2}
\label{formuleHC2}
\end{equation}
\noindent with $H_{c2//ab} \sim1.45\pm 0.05$ T and $H_{c2//c}\sim 0.85 \pm0.05$ T (see dotted line in Fig.3), i.e. corresponding to a mass anisotropy ratio of $\sim 1.7 \pm 0.1$ in good agreement with previous studies ~\cite{Morosan,Kacmarcik}.

\begin{figure}[h]
\begin{center}
\resizebox{8.5cm}{!}{\includegraphics{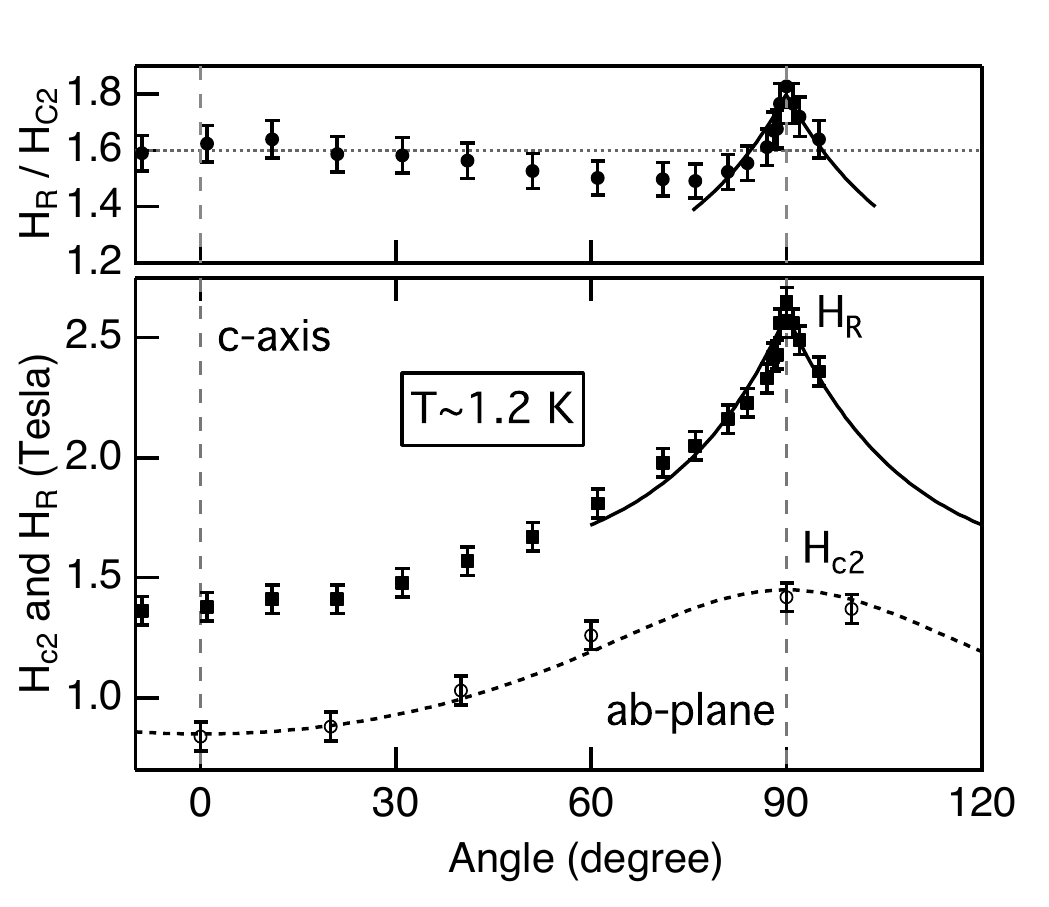}}
\caption{Angular dependence at $T=1.2$ K of the upper critical field $H_{c2}$ (open circles) and of the field $H_R$ (closed squares) on the bottom panel.  $H_{c2}$  is deduced from specific heat measurements (see Fig.2, upper panel). $H_R$ is deduced from resistivity measurements (see Fig.2, lower panel). The dotted line is a fit to the data using the standard London model for 3D anisotropic system (see equation \ref{formuleHC2} and text for details). The solid line corresponds to the angular dependence of the surface conductivity (see equation \ref{formuleHC3} and text for details). The solid line aim to highlight the characteristic cusp-like shape of surface superconductivity for magnetic fields close to the layer plane. Note that $H_R$ is surprisingly much larger than $H_{c2}$ for $\theta\rightarrow 0$. Upper panel: angular dependence of the $H_R/H_{c2}$ ratio showing that this ratio only weakly depends on the field orientation in striking contrast with standard surface superconductivity (solid line).} 
\label{fig3}
\end{center}
\end{figure}

On the contrary the shape of the resistivity transition strongly depends on the field orientation. Indeed, as shown in Fig.2 (lower panel), for $\theta\lesssim 60^\circ$ a sharp increase of the resistivity is observed for fields close to $H_{c2}$. Note that this onset field corresponds to the onset field of the specific heat transition showing that $R\rightarrow 0$ as soon as a significant part of the sample becomes superconducting and forms percolation paths shorting out the electrical resistance of the sample. A kink is then clearly visible in the $R$ versus $H$ curve and $R$ only reaches its normal state value for $H=H_R>>H_{c2}$ (open squares and short solid lines in Fig.2, lower panel). Finally, for $\theta \gtrsim 75^\circ$ the resistive transition becomes very broad and we did not observe anymore any clear signature of the upper critical field in the magnetoresistance data (see Fig.2 for $\theta = 90^\circ$). 

The angular dependence of $H_R$ is displayed in Fig.3. As shown, a sharp cusp is clearly visible close to $90^\circ$. This cusp is characteristic of standard surface conductivity and the angular dependence of $H_R$ ($\equiv H_{c3}$) can then be well fitted by the surface superconductivity formula proposed by K. Yawafuji {\it et al.} \cite{Yamafuji}:
\begin{equation}
\left[\frac{H_{c3}(\theta)}{H_{c3\|ab}}sin\theta(1+\frac{1-cos\theta}{2tan\theta})\right]^2+\frac{H_{c3}(\theta)}{H_{c3\|c}}cos\theta=1
\label{formuleHC3}
\end{equation}
\noindent with $ H_{c3\|c} \approx H_{c2\|ab}  \approx H_{c3\|ab}/1.7$ as suggested by D. Saint-James and P. G. De Gennes \cite{deGennes}. Note that the numerical value of the $H_{c3\|ab}/H_{c3\|c}$ ratio can actually significantly differ (both experimentally and theoretically) from the standard 1.7 Saint-James - De Gennes value (see [17] and discussion below). The fit is only intended to emphasize the cusp-like angular dependence of $H_R$ in the vicinity of the $ab$-planes.  Note also that this formula has been derived for isotropic superconductors and does not take into account the angular dependence of the upper critical field in anisotropic systems. However, this dependence can be neglected close to $90^\circ$ leading to a very reasonable fit to the data in this angular range. If this angular dependence can hence be reasonably attributed to {\it standard} surface effects close to the $ab-$plane, this is absolutely not the case for $\theta \rightarrow 0$. Indeed, in this latter case, surface effects would be expected to vanish for fields perpendicular to the planes and $H_R$ would hence be expected to tend towards $H_{c2}$. However, as shown on the upper panel of Fig.3, $H_R/H_{c2}$ surprisingly remains almost constant on the entire angle range emphasizing the existence of some unexpected superconductivity persisting well above the bulk upper critical field effects for all field directions. 

A change of slope in the magnetoresistance of CuTiSe$_2$ crystals prior to the establishment of bulk superconductivity has been previously reported by Morosan {\it el al.} \cite{Morosan}. This feature was attributed, with caution, to Cu  inhomogeneities, but it is worth noting that the superconducting transition is very sharp in zero magnetic field for both reference~\cite{Morosan} and our samples, rather attesting for the good homogeneity of the samples. A similar effect has also been reported in other layered systems such as NbSe$_2$ \cite{RH_NbSe2} (note that Fig.3 in this work is strikingly similar to Fig.2 of the present work) and MgB$_2$ \cite{Lyard}. In this latter system, this effect has been attributed to surface currents flowing on the lateral side faces due to the fact that the contact pads were covering the whole lateral surface of the platelet. This is however not the case here as point like contacts were made on the top of the platelet. Note that the observation of such an effect in MgB$_2$ tends also to rule out any explanation based on the presence of a charge order in the other compounds. 

\begin{figure}[h]
\begin{center}
\resizebox{8.5cm}{!}{\includegraphics{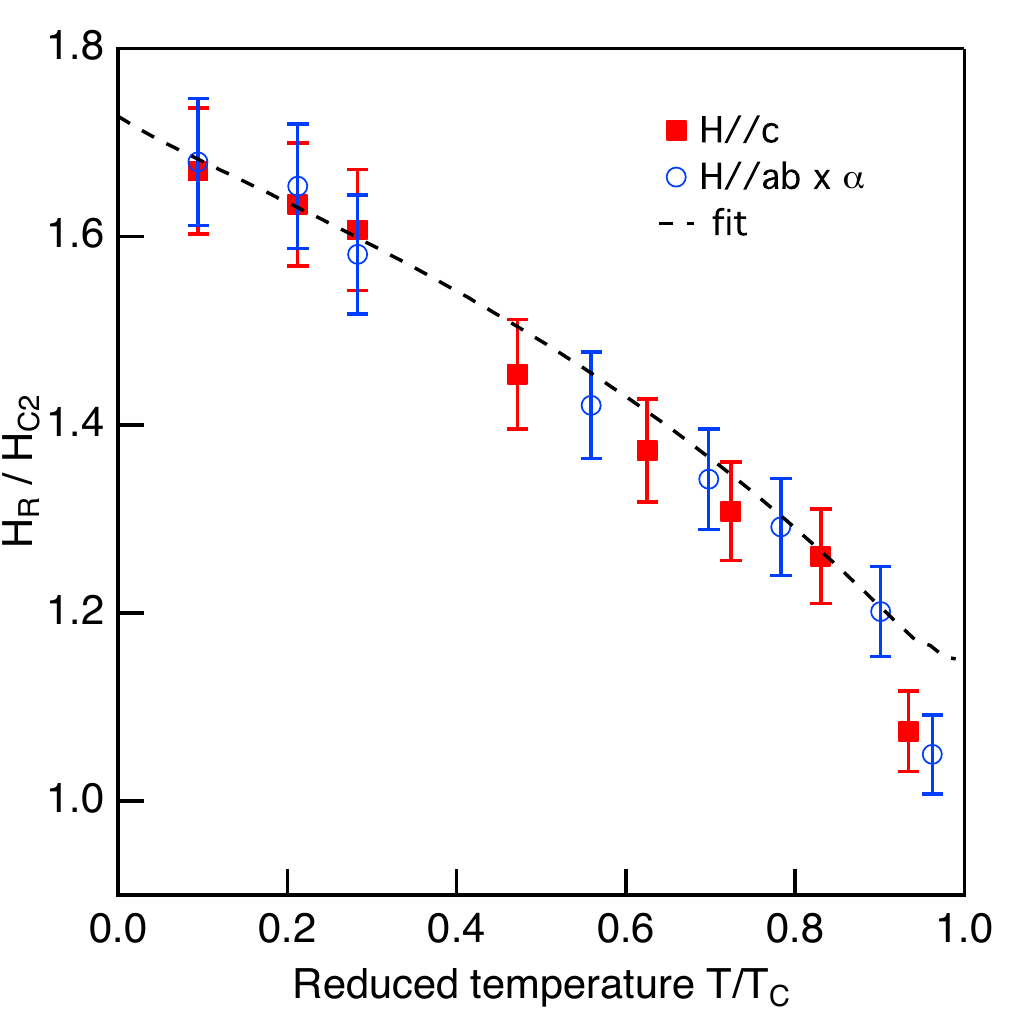}}
\caption{Temperature dependence of the $H_R/H_{c2}$ ratio for $H\|c$ and $H$ close to the ab-plane (see text and Fig.2 for definition of the fields). The dotted line is a fit to the data for $H\|ab$ assuming the presence of surface (or interface) layers with depressed superconducting properties leading to the existence of a boundary nucleation field, larger than $H_{c2}$ but with $H_R/H_{c2}$ significantly smaller than the standard 1.7 Saint James - de Gennes value (for $T \rightarrow T_c$). For fields close to the $ab$-planes, the curves can be shifted by a small normalization coefficient $\alpha\sim0.9-1$ depending on the exact orientation of the field compared to the planes but note that this temperature dependence is strikingly - and surprisingly - similar for both field orientations.}
\label{fig4}
\end{center}
\end{figure}

Figure~\ref{fig4} displays the temperature dependence of the $H_{R}$/$H_{c2}$ ratio for fields along the $c$-axis and close to the $ab$-plane. As raw specific heat and resistivity measurements were not taken at the same temperature points, the  $H_{c2}$ values were interpolated using a WHH formula~\cite{Werthamer,Gor'kov} (previous studies \cite{Morosan,Kacmarcik} demonstrated that such a WHH formula leads to a very reasonable fit to the data for both field-orientations). As pointed out above, for fields close to the $ab$-planes, $H_R$ can be attributed to standard surface superconductivity but, as shown, the $H_R/H_{c2}$ ratio decreases with temperature, hence being much smaller than the Saint James - de Gennes 1.7 value. However this value has been derived in the case of an interface between the superconducting materials and vacuum \cite{RemValue}  and much smaller values are actually expected for $T\rightarrow T_c$ in presence of a top layer with depressed superconducting properties \cite{Hu_twoL}. 

Assuming that the system consists of two (semi-infinite) surface (S) and bulk (B) superconductors in contact ($T_c^S<T_c^B$) one can solve the linearized Ginzburg-Landau equations in presence of a magnetic field parallel to the interface \cite{Fink} in order to calculate the {\it surface} critical field $H_{c3}$ which is defined by the boundary conditions at the interface between the two sub-layers. $H_{c3}/H_{c2}$ is then given by a numerical function $F(\gamma,\alpha(T))$ depending on $\gamma=\sigma_N^B/\sigma_N^S\sim (m_S/m_B)\times(n_B/n_S)$ and $\alpha(T) = [\xi_S/\xi_B]^2$ where $\sigma_N"$, $m_i$, $n_i$ and $\xi_i$ are the normal state conductivity, effective masses, carrier densities and coherence lengths in the bulk ($i=B$) and at the surface ($i=S$), respectively.  Very reasonable fit to the data (dotted line in Fig.4) can be obtained assuming that the surface critical temperature $T_c^S$ is on the order of $T_c^B/\gamma$ and the zero temperature surface coherence length ratio $\xi^0_S/\xi^0_B \sim 0.74-\gamma/130$.  

An estimation of $\gamma$ from other techniques is then necessary in order to be conclusive on the surface superconducting properties.  Note that band structure calculations of TiSe$_2$ slabs predict a small 10$\%$ decrease of the surface Ti 3d bandwidth with respect to the bulk \cite{Fang}, and so a similar small increase of the effective mass at the surface. Besides, the Fermi energy and velocity measured by ARPES in Cu-doped TiSe$_2$ allow to predict a coherence length very close to the one obtained from bulk $H_{c2}$ measurements \cite{Qian}. As the mean free path of photoelectrons is only a few $\AA$, it is clear that the electronic structure of the surface is very similar to the bulk electronic structure. In addition, a change of Cu-doping up to 40\% close to our optimal doping only leads to a change by a factor 2 of the conductivity~\cite{Morosan2006}. As the crystal growth has been made at equilibrium, higher Cu-doping variation are very unlikely and in our case the surface and the bulk should not be seen as two different materials, but rather as internal modifications of the same superconductor, thus $\gamma$ is reasonably expected to be of the order of unity.   

It is worth noting that a very strong lock-in of the vortices along the $ab-$planes has been observed recently \cite{Medvecka} in Hall probe magnetization measurements. This lock-in effect clearly indicates a strong modulation of the core energy (i.e. of the order parameter) along the $c-$direction suggesting the existence of "dead"-layers ($\|ab$) with depressed superconducting properties. Those layers could account for the observation of our surface (or interface) superconductivity for $H\|ab$. {\it Surface} would then more refer to a boundary nucleation field (at the interface between the bulk and the "dead"-layers) being larger than the bulk upper critical field \cite{Fink}. Even if both the angular and temperature dependence of $H_R$ could then be well described by the standard theory for field orientations close to the $ab-$planes, the origin of surface and/or interface effects for $H\|c$ remains to be clarified. 

\section{Conclusion}
In summary, we have demonstrated through the comparison between specific heat and resistivity measurements the existence of surface (or interface) superconductivity in Cu$_{10\%}$TiSe$_2$ single crystals. Surprisingly, this surface superconductivity is present for all  magnetic field orientations either parallel or perpendicular to the layers. For $H$ close to the $ab$-planes, the temperature dependence of the $H_R/H_{c2}$ ratio indicates the presence of surface (or interface) layers with depressed superconducting properties. The influence of those layers on the superconducting properties for $H\|c$ deserves further theoretical and experimental works. 

\section*{Acknowledgments}
This work is supported by the French National Research Agency through Grant No. ANR-12-JS04-0003-01 SUBRISSYME. JK has been supported by APVV-0036-11 and APVV-0405-14. 

\section*{References}
\bibliographystyle{unsrt}





\end{document}